\documentclass{article}
\usepackage{spconf,amsmath,graphicx, amssymb}
\usepackage{multirow,ragged2e, array}
\usepackage{flushend}   
\usepackage{stfloats}
\usepackage{enumitem}
\setlist{nosep, leftmargin=14pt}
\newcommand\blfootnote[1]{%
  \begingroup
  \renewcommand\thefootnote{}\footnote{#1}%
  \addtocounter{footnote}{-1}%
  \endgroup
}
\usepackage{mwe} 

\title{Deep Image Priors for Magnetic Resonance Fingerprinting with pretrained Bloch-consistent denoising autoencoders}

\name{%
\begin{tabular}{@{}c@{}}
Perla Mayo \quad 
Matteo Cencini \quad 
Ketan Fatania \quad
Carolin M. Pirkl \quad 
Marion I. Menzel \\
Bjoern H. Menze \quad
Michela Tosetti \quad
Mohammad Golbabaee
\end{tabular}}
\address{\footnotesize}
\begin{document}
\ninept
\maketitle
%
\begin{abstract}
The estimation of multi-parametric quantitative maps from Magnetic Resonance Fingerprinting (MRF) compressed sampled acquisitions, albeit successful, remains a challenge due to the high underspampling rate and artifacts naturally occuring during image reconstruction. Whilst state-of-the-art DL methods can successfully address the task, to fully exploit their capabilities they often require training on a paired dataset, in an area where ground truth is seldom available. In this work, we propose a method that combines a deep image prior (DIP) module that, without ground truth and in conjunction with a Bloch consistency enforcing autoencoder, can tackle the problem, resulting in a method faster and of equivalent or better accuracy than DIP-MRF.
\end{abstract}
\blfootnote{PM and MG are with the University of Bristol, UK; MC is with the INFN Pisa division, Italy; KF is with the University of Bath, UK; CMP and MIM are with GE HealthCare, Germany; MIM is also with the Technische Hochschule Ingolstadt, Germany; BHM is with the University of Zurich, Switzerland; and MT is with the IRCCS Stella Maris, Italy}
\begin{keywords}
magnetic resonance fingerprinting, deep learning, deep image priors, quantitative magnetic resonance imaging
\end{keywords}
%

\section{Introduction}
\label{sec:intro}
Traditional Magnetic Resonance Imaging (MRI) provides weighted images, where image contrast is determined by tissue properties and scan parameters. While useful, weighted images prevent objective and reproducible assessments. Quantitative imaging, particularly multiparametric MRI, can bridge this gap. In Magnetic resonance fingerprinting (MRF)\cite{ma2013mrf} signals are acquired in a transient rather than a steady state. That is, the image series are acquired while the signal evolves over time. This, combined with spatial undersampling can significantly speed up the scanning process. 
A pattern matching algorithm is used to determine the tissue parameters. Acquiring compressed measurements enables the fast encoding of multiple tissue parameters - usually T1 and T2 relaxation times - in a single time-efficient scan. However, the gain in speed provided by the high undersampling comes to the cost of introducing aliasing artifacts present in the reconstructed images.

This issue requires addressing through advanced quantitative image reconstruction techniques. While state-of-the-art supervised deep learning methods\cite{fang2019supervisedmrf} can help, obtaining artifact-free "ground truth" quantitative maps remains a challenge\cite{fatania2023nonlinearequivariant} as their acquisition requires dense sampling long-scanning times, which are impractical. Deep Image Prior (DIP)\cite{ulyanov2018dip} has emerged as a successful approach for ground-truth-free image reconstruction with applications in medical imaging\cite{gong2018petdip}, compressed sensing\cite{yoo2021dipdynamicmri}, and more recently, MRF\cite{hamilton2022dipmrf}. DIP's strength lies in the well-crafted architecture of convolutional neural networks that can sufficiently act as an implicit image prior without the need for pretraining them on a dataset. In \cite{hamilton2022dipmrf}, two networks are trained in parallel, the first one (a Unet\cite{ronneberger2015unet}) is trained solely on k-space data to produce clean fingerprints, which are then inputted to the second model (a fully connected network) in charge of \textit{encoding} these fingerprints into parameter maps. The training of the encoding module is aided by a network pre-trained on a dictionary of signatures that \textit{decodes} the parameter maps back to the fingerprints. Thus, the objective of the encoding network during training is to reduce the error between the Unet and the decoder outputs. The result of this is an architecture capable of estimating quantitative parametric maps that does not require ground truth.

However, the current DIP-MRF algorithm\cite{hamilton2022dipmrf} is computationally intensive, requiring a large number of iterations with slow and sometimes unstable convergence. To address this, we introduce Bloch Autoencoder Regularized Deep Image Prior (BARDIP), an enhanced DIP reconstruction method capable of retrieving multi-parametric maps using only the frequency domain data. BARDIP demonstrates up to 30x faster convergence than the current DIP-based baseline on simulated data while rivalling quantitative mapping accuracy.

\section{Methods}
\label{sec:methods}
BARDIP iteratively estimates the quantitative maps $\textbf{q} =$ \textit{\{T1, T2, Proton Density (PD)\}} and corresponding fingerprints, formally referred as time series of magnetisation images (TSMI), $\textbf{x} \in \mathbb{C}^{N \times K}$ for $N$ pixels across $K$ Singular Value Decomposition (SVD) channels, using undersampled k-space data $\textbf{y} \in \mathbb{C}^{cM \times L}$ for $M$ samples from $c$ coils and $L$ timeframes, obtained from one MRF scan. It solves the following inverse problem: 
\begin{equation}
    \label{eq:inverse_problem}
    \textbf{y} \approx A(\textbf{x}), \;\;
    s.t.\;\;\textbf{x}_p=\text{PD}_p \cdot B(\text{T1}_p,\text{T2}_p), \;\forall p : pixels, 
\end{equation}

\noindent
where $A$ is the linear forward/acquisition operator, comprising nonuniform-FFT, coil sensitives and SVD reduction\cite{mcgivney2014svdmrf}; and $B$ is the nonlinear Bloch response, relating quantitative maps to TSMI voxel-wise.

\begin{figure*}[t]

    \begin{minipage}[b]{.65\linewidth}
        \centering
        \centerline{\includegraphics[width=10cm]{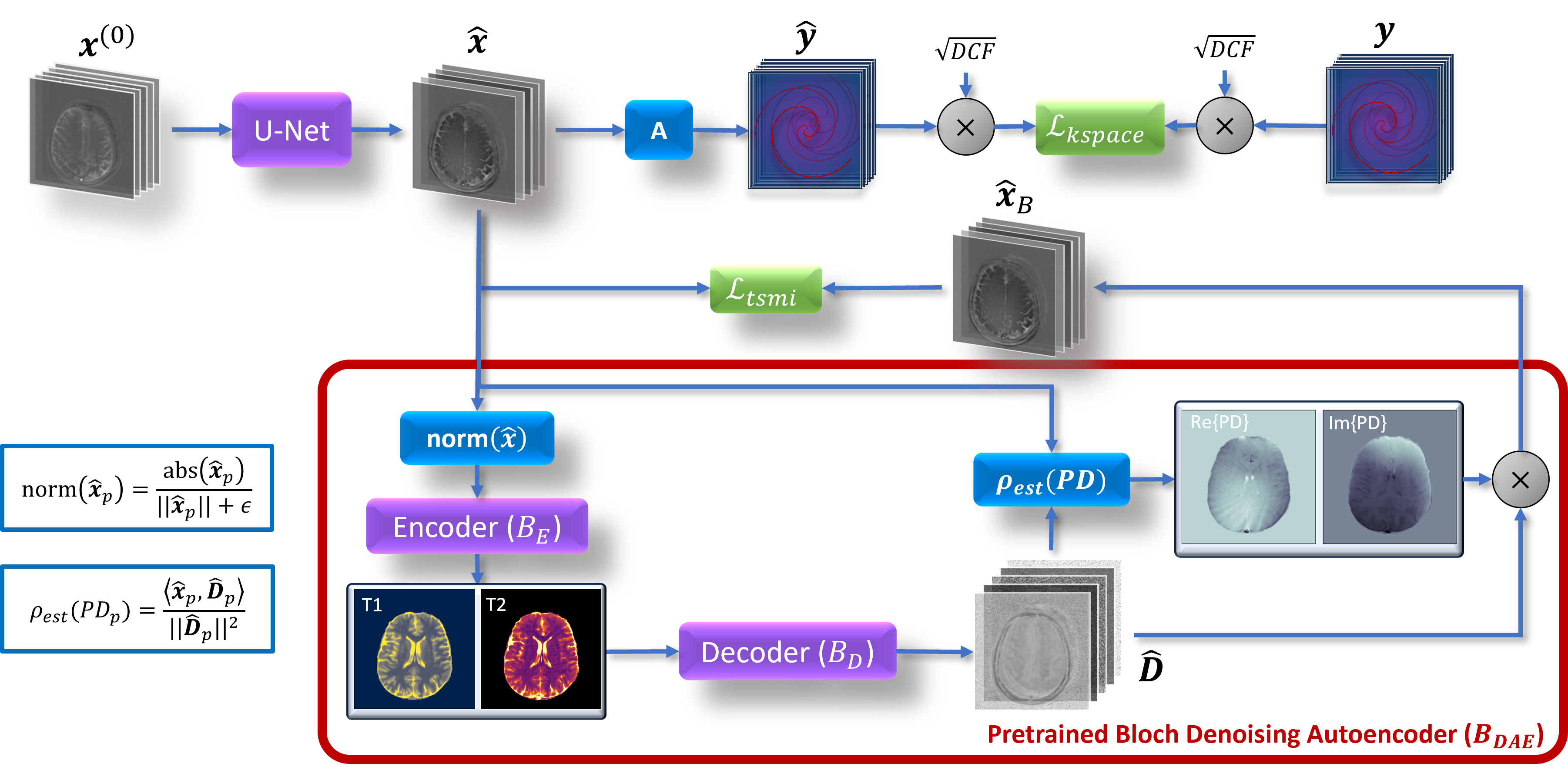}}
        \centerline{(a)}\medskip
    \end{minipage}
    \hfill
    \begin{minipage}[b]{0.3\linewidth}
        \centering
        \centerline{\includegraphics[width=5.5cm]{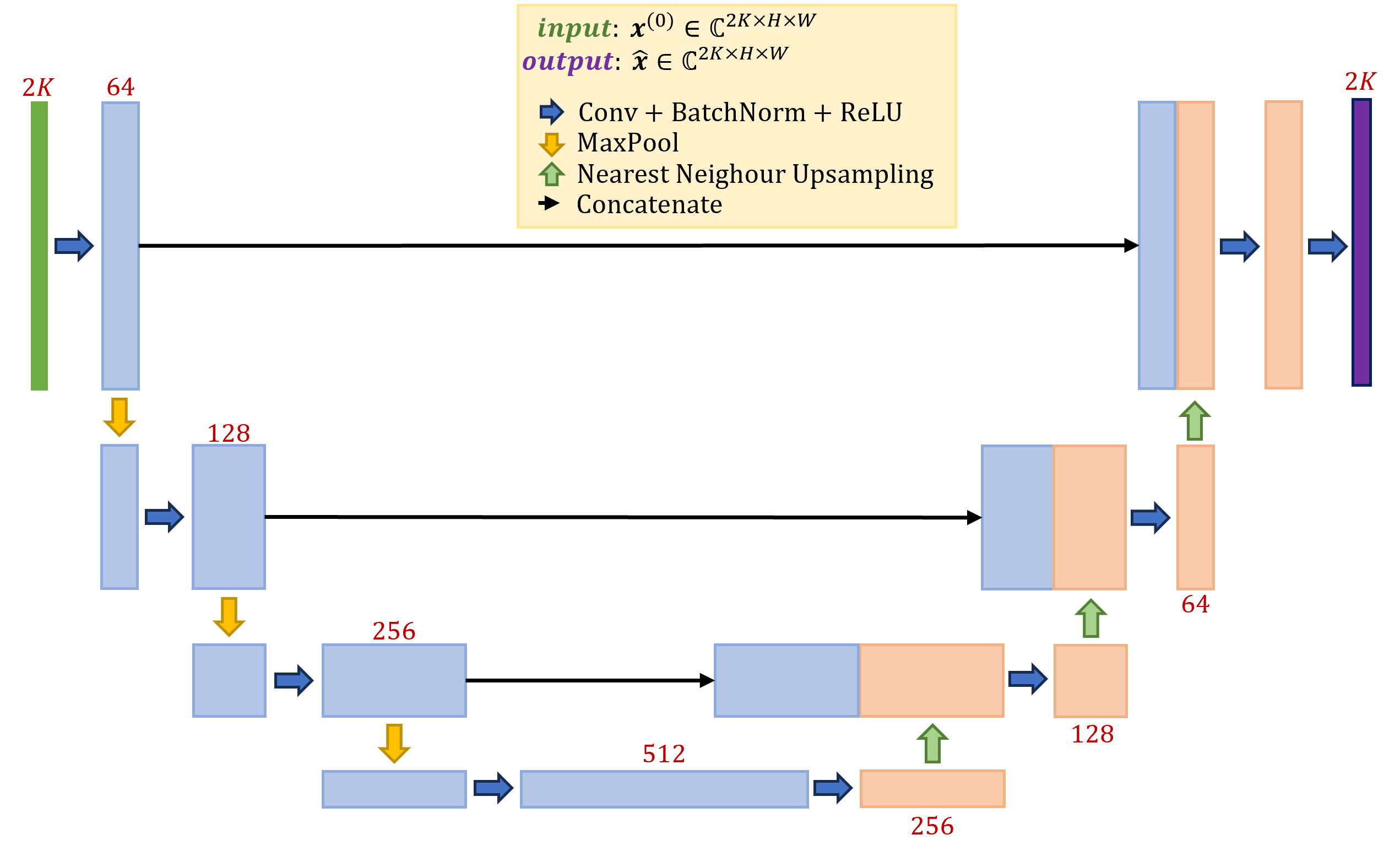}}
        \vspace{0.5cm}
        \centerline{(b)}\medskip
    \end{minipage}
    \caption{\footnotesize (a) Proposed pipeline composed of three neural networks, two of them pretrained. The computation of the forward operator as well as the use of the $\textbf{B}_{\textbf{DAE}}$ module enforce the consistency with the data acquisition model and the Bloch equations. (b) Unet used for the experiments reported.}
    \label{fig:pipeline}
\end{figure*}

\subsection{BARDIP Overview}
An overview diagram of BARDIP is given in Fig.  \ref{fig:pipeline} (a). This approach employs the following three major components.
\newline\newline
\textbf{Unet}\cite{ronneberger2015unet}: Acts as a DIP prior, working to de-alias the input TSMI obtained from scaled back-projection:

\begin{equation}
    \textbf{x}^{(0)} \stackrel{def}{=} \frac{||\textbf{y}||}{||AA^H \textbf{y}||}A^H(\textbf{y}),
    \label{eq:x_0}
\end{equation} 

\noindent
where $A^{H}$ is the hermitian transpose of $A$, to output the cleaner version  $\hat{\textbf{x}}\stackrel{def}{=}\textbf{Unet}_\Theta (\textbf{x}^{(0)})$, with $\textbf{x}^{(0)}, \hat{\textbf{x}}  \in \mathbb{C}^{N \times K}$.
\newline
\newline
\textbf{Bloch Denoising Autoencoder} ($\textbf{B}_{\textbf{DAE}}$):
It is comprised of two fully connected neural networks for quantitative mapping, nonlinear dimensionality-reduction, and outputting Bloch-consistent denoised TSMI $\hat{\textbf{x}}_B$. The networks are architecturally similar with two hidden layers of 300 neurons each, however, each aiming to achieve a different objective and processing different inputs/outputs:

\begin{itemize}
    \item The \textbf{Encoder} ($\textbf{B}_\textbf{E}$) acts as an inverse of $B$ to pixel-wise project a noisy TSMI input to T1 and T2 parameters\cite{cohen2018drone, golbabaee2019geometry}. Taking as input the $\ell_2$ normalised TSMI and producing as output estimation maps for T1 and T2.
    \item The \textbf{Decoder} ($\textbf{B}_\textbf{D}$): a.k.a. Blochnet \cite{chen2020blochnet}, approximates $B$ for mapping T1/T2 to a Bloch-consistent denoised TSMI. It receives the T1/T2 output of the encoder and outputs the scaled TSMI $\hat{\textbf{D}} \in \mathbb{C}^{N \times K}$. 
\end{itemize}
Our approach estimates the complex PD analytically~\cite{davies2014blip} by computing $\text{PD}_p=<\hat{\textbf{x}}_p, \hat{\textbf{D}}_p>/{||\hat{\textbf{D}}_p||^2}$.
And thus, $\hat{\textbf{x}}_{B}$ is obtained from the pixel-wise multiplication of $\hat{\textbf{D}}$ and PD.
\newline
\newline
\textbf{Multitasking/coupled loss:}
Optimises the $\textbf{Unet}$ parameters $\Theta$, hence, to reconstruct $\hat{\textbf{x}}$ (note that the pretrained $\textbf{B}_{\textbf{DAE}}$ is frozen here):
 \begin{equation}
    \label{eq:loss}
    \begin{split}
        \mathcal{L}   & = \mathcal{L}_{kspace} + \lambda \mathcal{L}_{TSMI} \\ 
            & = ||\sqrt{DCF}\cdot \textbf{y}- \sqrt{DCF} \cdot A(\hat{\textbf{x}})||_2^2 + \lambda ||\hat{\textbf{x}} - \hat{\textbf{x}}_{B}||_2^2 
    \end{split}
\end{equation}

The square root of the density compensation function (DCF) is applied as a preconditioner within $\mathcal{L}_{kspace}$ to accelerate optimization. It also utilises the weight term $\lambda = 1e^{-5}$ to prevent either loss dominating the learning. Besides k-space data consistency, eq. \eqref{eq:loss} uses two additional priors for reconstruction: 1) DIP spatial image prior via $\textbf{Unet}$, and 2) Bloch-consistency via $\mathcal{L}_{TSMI}$ and $\textbf{B}_{\textbf{DAE}}$. DIP-MRF only used the k-space consistency loss for optimizing the $\textbf{Unet}$.

\subsection{Training Models}
$\textbf{Unet}$ optimization was performed using the ADAM optimizer with a learning rate of 1e-4, on the top 5 SVD basis of both approaches over 30k epochs. This selection of hyperparameters aligns with the early stopping strategy reported by DIP-MRF, aimed at preventing overfitting to undesired artifacts. Real and imaginary parts of complex arrays were formatted whenever needed such that they conform two different channels, therefore, instead of using K channels of complex numbers, the networks receive/output 2K channels of real values.

The $\textbf{B}_{\textbf{DAE}}$ module was pretrained in a supervised manner for 1k epochs on an SVD-MRF\cite{mcgivney2014svdmrf} dictionary, using pairs of T1/T2 values and their Bloch responses (fingerprints) obtained from Extend-Phase-Graph simulations\cite{weigel2015epg}. Training involved multiplying fingerprints by random complex phasors, adding complex Gaussian noise ($\sigma = 0.01$), and minimising the sum of MSE loss between the denoised and noiseless complex fingerprints, and MAE losses between predicted and true (T1,T2) values, as described in eq. \eqref{eq:bdae_loss}.

\begin{equation}
    \label{eq:bdae_loss}
    \mathcal{L}_{E}=||\textbf{q}_{T1}- \hat{\textbf{q}}_{T1}||_1 + 10\cdot || \textbf{q}_{T2} - \hat{\textbf{q}}_{T2}||_1 + \lambda_{E}||\textbf{x}-\hat{\textbf{x}}||_2^2
\end{equation}

The value of the weight term in eq. \eqref{eq:bdae_loss}, $\lambda_{E}$, was empirically determined to be $\lambda_E=0.1$. Note, while DIP-MRF has architecturally similar encoder/decoder modules, only the decoder $\textbf{B}_\textbf{D}$ is pretrained using the available MRF dictionary, whereas the encoder $\textbf{B}_\textbf{E}$ is self-supervisedly trained in parallel to the $\textbf{Unet}$.

\section{Numerical Experiments}
\label{sec:results_and_discussion}
Methods were tested on simulated and healthy volunteer brain axial slices, using Steady State Precession (FISP) sequence flip angle schedule~\cite{jiang2015fisp}, TR/TE/TI = 10/1.908/18 ms, $L$=1000 repetitions (timeframes), variable-density spiral readouts, $N=230\times230$ matrix size, 1mm in-plane resolution and 5mm slice thickness. In-vivo MRF data was acquired on 3T GE scanner (MR750w system GE Healthcare, Waukesha, WI) with 8-channel receive-only head RF coil. Acquisition parameters were also used to simulate additional single-coil MRF data using "ground-truth" qmaps from MAGiC scans of 17 brain slices/3 volunteers with added Gaussian noise (SNR=35 and 40 dB) to simulated k-space measurements. 
Simulated data (with ground-truth) was used to quantitatively assess the methods, whereas in-vivo data (without ground-truth) was used for qualitative assessment.

We implemented the DIP-MRF\cite{hamilton2022dipmrf} baseline according to the paper’s specifications, but using our custom $\textbf{Unet}$ architecture (Fig. \ref{fig:pipeline} (b)) with LR=$1e^{-4}$, which demonstrated better results for our dataset. Other training parameters remained as reported.

\subsection{Results}
\begin{figure}[tb]
    \begin{minipage}[b]{\linewidth}
        \centering
        \centerline{\includegraphics[width=9cm]{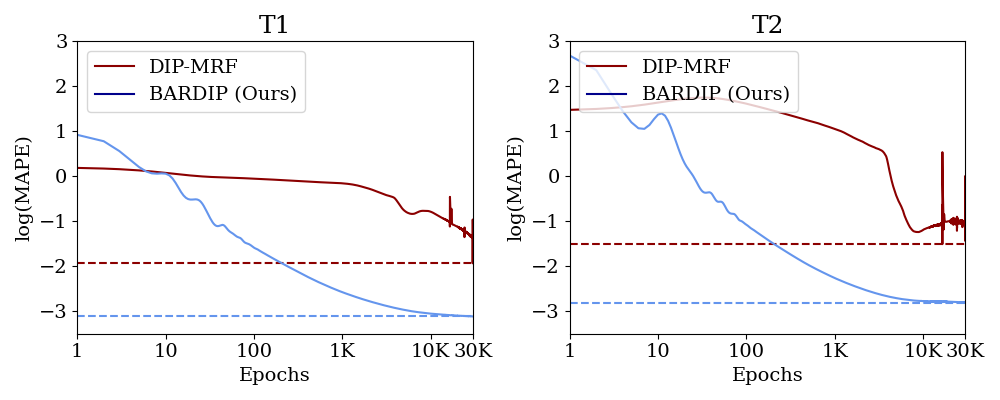}}
    \end{minipage}
    \begin{minipage}[b]{\linewidth}
        \centering
        \centerline{\includegraphics[width=9cm]{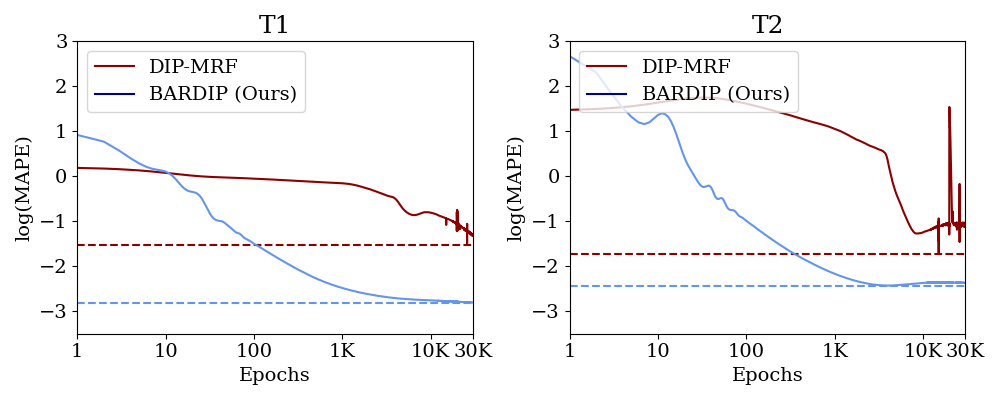}}
    \end{minipage}
    \caption{\footnotesize T1/T2 reconstruction errors vs iterations for learned approaches on simulated data. Dashed lines indicate the lowest value found across all iterations for the respective approach. Top: SNR 40. Bottom: SNR 35.} 
    \label{fig:loglogmape}
\end{figure}

Fig. \ref{fig:loglogmape} shows the decreasing rate of Mean Average Percentage Error (MAPE) of T1/T2 parameters for both techniques during the reconstruction of a single slice from the simulated data for SNR 35 and SNR 40. Note that these metrics are purely for reporting purposes and are not used during training. Table ~\ref{tab:metrics_simulated} reports the performance metrics on skull-stripped simulated brain maps of the reconstructed quantitative maps at 1k and 30k iterations, averaged across the 17 slices. 
Lastly, Figure \ref{fig:all_maps_invivo} showcases the reconstructed q-maps for the real scan slice after iterations 1k and 30k for the two approaches and with two acquisition settings: L=1000 (original scans) and L=500 (retrospectively truncated scans).

\begin{table}[t]
\begin{center}
    \scriptsize
    \begin{tabular}{ | m{1cm} | m{0.4cm} || m{0.6cm}  | m{0.6cm} |  m{0.6cm} || m{0.6cm}  | m{0.6cm} |  m{0.6cm} |} 
         \hline
         \multirow{2}{1cm}{\Centering \textbf{Approach}} & \multirow{2}{0.4cm}{ \Centering \textbf{Iter.} } & \multicolumn{3}{c||}{\textbf{SNR 40}} & \multicolumn{3}{ c|}{\textbf{SNR 35}} \\ \cline{3-8}
                & & \Centering \textbf{T1} & \Centering \textbf{T2} & \Centering \textbf{PD} & \Centering \textbf{T1} & \Centering \textbf{T2} & \Centering \textbf{PD} \\
         \hline\hline 
         \multirow{2}{1cm}{\Centering \textbf{DIP-MRF}} 
                & \Centering 1k   & \Centering  80.59 & \Centering 259.7 & \Centering 10.80
                                  & \Centering  80.57 & \Centering 259.7 & \Centering 10.79 \\
                & \Centering 30k  & \Centering  21.92 & \Centering 35.96 & \Centering 18.10
                                  & \Centering  21.89 & \Centering 31.35 & \Centering 17.01  \\ 
        \hline
        \multirow{2}{1cm}{\Centering \textbf{BARDIP (Ours)}} 
                & \Centering  1k  & \Centering 7.33 & \Centering 10.35 & \Centering 21.31
                                  & \Centering 8.08 & \Centering 11.24 & \Centering 22.98 \\ 
                & \Centering 30k  & \Centering \textbf{4.22} & \Centering \textbf{5.47} & \Centering \textbf{31.21} 
                                  & \Centering \textbf{5.56} & \Centering \textbf{8.29} & \Centering \textbf{28.83} \\
                                  
        \hline
    \end{tabular}
    \caption{\footnotesize The reconstruction errors, measured as Mean Absolute Percentage Error (MAPE) for T1 and T2 maps (in \%), and Peak Signal-to-Noise Ratios (PSNRs) for PD (in dB), are reported. These metrics were averaged across simulated data comprising 17 brain slices from 3 subjects. The evaluated techniques include DIP-MRF and BARDIP, each assessed at 1,000 and 30,000 iterations. The best overall metrics are highlighted in bold.}
    \label{tab:metrics_simulated}
\end{center}
\end{table}

\subsection{Discussion}
\begin{figure*}[tbh]
    \centering
    \centerline{\includegraphics[width=18cm]{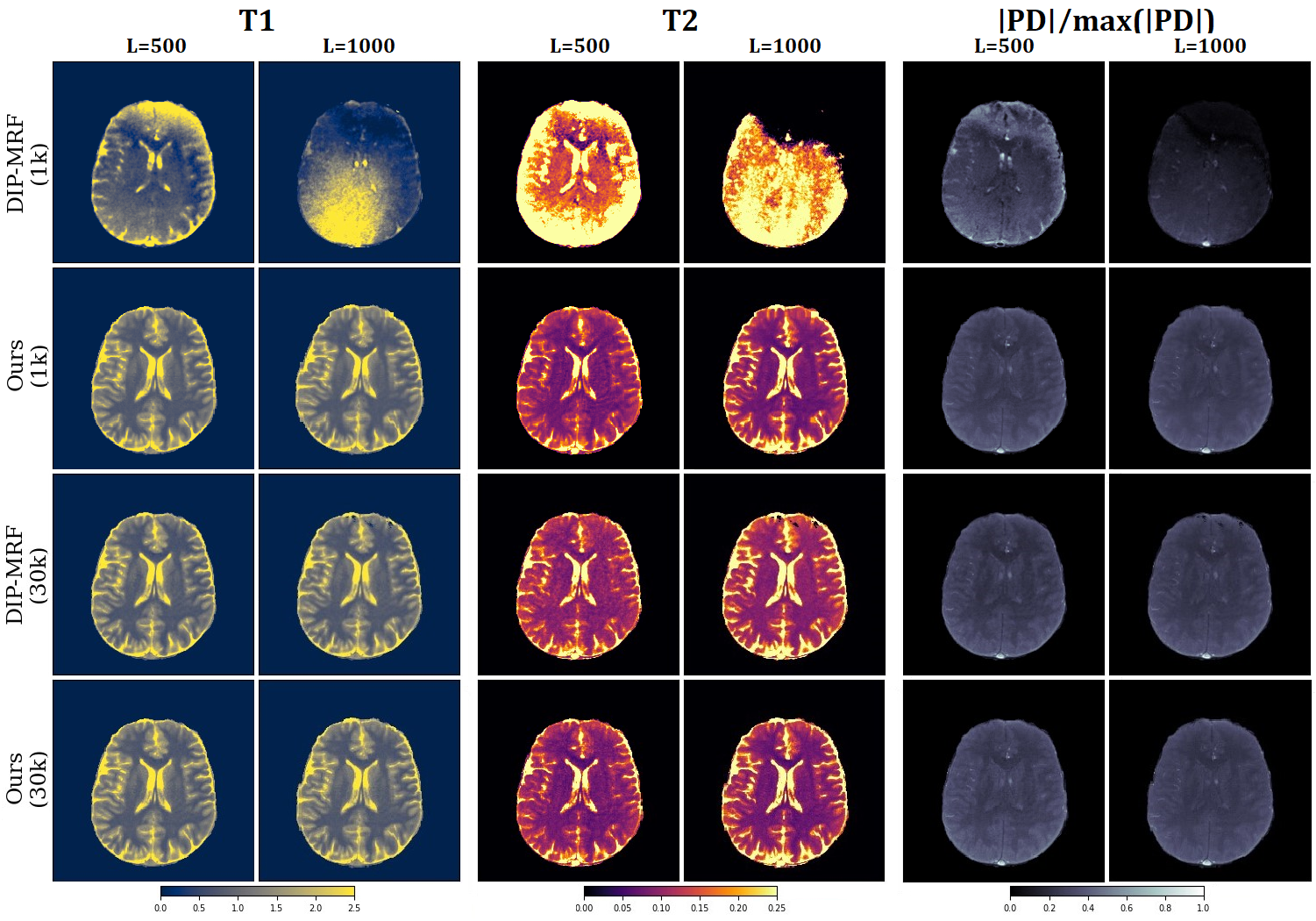}}
    \caption{\small Tissue maps of the estimations of both approaches on the single slice of real data assessed with L=1000 and L=500. T1 and T2 are in seconds.}
    \label{fig:all_maps_invivo}
\end{figure*}

The appeal of ground-truth free approaches such as those based on DIP architectures can be counterbalanced by the need of lengthy iterations and the early stopping required to prevent overfitting to the corrupted data. This issue becomes more prevalent when the techniques become unstable, as seen in the baseline approach in Fig. \ref{fig:loglogmape}. As such, there is no guarantee that at the stopping point the reconstructed image will lie away from the shown outliers (jumps in T1/T2 MAPEs). Tuning the LR can help to reach steadiness at the expense of longer training. In comparison, BARDIP exhibits a steady and more rapid decrease of MAPE during the entire course of training, reaching an acceptable error level within 1k iterations and plateauing after 10k, which suggests an even earlier stop would be acceptable for this approach. This is shown quantitatively in both in Fig. \ref{fig:loglogmape} and Table \ref{tab:metrics_simulated}, and qualitatively in Fig. \ref{fig:all_maps_invivo}. This gain in computation can be attributed to various factors: a) the pretraining of the $\textbf{B}_{\textbf{DAE}}$ module on the available MRF dictionary, as proposed in this work, which differs from DIP-MRF, where the encoder is self-supervisedly trained during reconstruction iterations, b) the choice of $\hat{\textbf{x}}^{(0)}$, which in this work corresponds to the scaled back projection defined in Eq. \eqref{eq:x_0}. Further, as iterations continue, BARDIP’s accuracy improves, without reaching the undesirable overfitting. We attribute this performance to the additional regularisation that BARDIP utilises for solving the problem through its Bloch-consistency enforcing coupled-loss. 

\section{Conclusion}
\label{sec:conclusion}
In this paper, we introduced BARDIP, a faster MRF reconstruction method based on deep image priors, with comparable quantitative mapping accuracy than baseline. We showed that exploiting the knowledge on the image formation and image modality aids in the estimation of better quantitative parametric maps. The proposed method benefits from utilising only the undersampled k-space measurements to produce parametric maps, i.e., a ground truth-free deep image reconstruction algorithm whilst also exhibiting stability for longer iterations.

\section{Compliance with Ethical Standards}
\label{sec:ethical}
This research study was conducted retrospectively using anonymised human subject scans made available by GE Healthcare who obtained informed consent in compliance with the German Act on Medical Devices.

\section{Acknowledgments}
\label{sec:acknowledgments}
This work has been carried out under the EPSRC grant EP/X001091/1. The extensive experiments were ran on the High Performance Computing (HPC) resources from the University of Bristol. 

\bibliographystyle{IEEEbib}
\bibliography{references_abbr}

\begin{thebibliography}{10}

\bibitem{ma2013mrf}
D.~Ma et~al.,
\newblock ``Magnetic resonance fingerprinting,''
\newblock {\em Nature}, vol. 495, no. 7440, pp. 187--192, 2013.

\bibitem{fang2019supervisedmrf}
Z.~Fang et~al.,
\newblock ``Deep learning for fast and spatially constrained tissue quantification from highly accelerated data in magnetic resonance fingerprinting,''
\newblock {\em IEEE transactions on medical imaging}, vol. 38, no. 10, pp. 2364--2374, 2019.

\bibitem{fatania2023nonlinearequivariant}
K.~Fatania, K.~Y Chau, C.M. Pirkl, M.I. Menzel, and M.~Golbabaee,
\newblock ``Nonlinear equivariant imaging: Learning multi-parametric tissue mapping without ground truth for compressive quantitative mri,''
\newblock in {\em 2023 IEEE 20th International Symposium on Biomedical Imaging (ISBI)}. IEEE, 2023, pp. 1--4.

\bibitem{ulyanov2018dip}
D.~Ulyanov, A.~Vedaldi, and V.~Lempitsky,
\newblock ``Deep image prior,''
\newblock in {\em Proceedings of the IEEE conference on computer vision and pattern recognition}, 2018, pp. 9446--9454.

\bibitem{gong2018petdip}
K.~Gong, C.~Catana, J.~Qi, and Q.~Li,
\newblock ``Pet image reconstruction using deep image prior,''
\newblock {\em IEEE transactions on medical imaging}, vol. 38, no. 7, pp. 1655--1665, 2018.

\bibitem{yoo2021dipdynamicmri}
J.~Yoo et~al.,
\newblock ``Time-dependent deep image prior for dynamic mri,''
\newblock {\em IEEE Transactions on Medical Imaging}, vol. 40, no. 12, pp. 3337--3348, 2021.

\bibitem{hamilton2022dipmrf}
J.I. Hamilton,
\newblock ``A self-supervised deep learning reconstruction for shortening the breathhold and acquisition window in cardiac magnetic resonance fingerprinting,''
\newblock {\em Frontiers in Cardiovascular Medicine}, vol. 9, pp. 928546, 2022.

\bibitem{ronneberger2015unet}
O.~Ronneberger, P.~Fischer, and T.~Brox,
\newblock ``U-net: Convolutional networks for biomedical image segmentation,''
\newblock in {\em Medical Image Computing and Computer-Assisted Intervention--MICCAI 2015: 18th International Conference, Munich, Germany, October 5-9, 2015, Proceedings, Part III 18}. Springer, 2015, pp. 234--241.

\bibitem{mcgivney2014svdmrf}
D.F. McGivney et~al.,
\newblock ``Svd compression for magnetic resonance fingerprinting in the time domain,''
\newblock {\em IEEE transactions on medical imaging}, vol. 33, no. 12, pp. 2311--2322, 2014.

\bibitem{cohen2018drone}
O.~Cohen, B.~Zhu, and M.S. Rosen,
\newblock ``Mr fingerprinting deep reconstruction network (drone),''
\newblock {\em Magnetic resonance in medicine}, vol. 80, no. 3, pp. 885--894, 2018.

\bibitem{golbabaee2019geometry}
M.~Golbabaee, D.~Chen, P.A. G{\'o}mez, M.I. Menzel, and M.E. Davies,
\newblock ``Geometry of deep learning for magnetic resonance fingerprinting,''
\newblock in {\em ICASSP 2019-2019 IEEE International Conference on Acoustics, Speech and Signal Processing (ICASSP)}. IEEE, 2019, pp. 7825--7829.

\bibitem{chen2020blochnet}
D.~Chen, M.E. Davies, and M.~Golbabaee,
\newblock ``Compressive mr fingerprinting reconstruction with neural proximal gradient iterations,''
\newblock in {\em Medical Image Computing and Computer Assisted Intervention--MICCAI 2020: 23rd International Conference, Lima, Peru, October 4--8, 2020, Proceedings, Part II 23}. Springer, 2020, pp. 13--22.

\bibitem{davies2014blip}
M.~Davies, G.~Puy, P.~Vandergheynst, and Y.~Wiaux,
\newblock ``A compressed sensing framework for magnetic resonance fingerprinting,''
\newblock {\em Siam journal on imaging sciences}, vol. 7, no. 4, pp. 2623--2656, 2014.

\bibitem{weigel2015epg}
M.~Weigel,
\newblock ``Extended phase graphs: dephasing, rf pulses, and echoes-pure and simple,''
\newblock {\em Journal of Magnetic Resonance Imaging}, vol. 41, no. 2, pp. 266--295, 2015.

\bibitem{jiang2015fisp}
Y.~Jiang, D.~Ma, N.~Seiberlich, V.~Gulani, and M.A. Griswold,
\newblock ``Mr fingerprinting using fast imaging with steady state precession (fisp) with spiral readout,''
\newblock {\em Magnetic resonance in medicine}, vol. 74, no. 6, pp. 1621--1631, 2015.

\end{thebibliography}

\end{document}